\documentclass[aps,pra,reprint,superscriptaddress,nofootinbib]{revtex4-1}
\usepackage{amsmath}
\usepackage{amssymb}
\usepackage{amsfonts}
\usepackage{mathrsfs}
\usepackage{graphicx}
\usepackage[caption=false]{subfig}
\usepackage{enumitem}
\usepackage[normalem]{ulem}
\usepackage[percent]{overpic}
\usepackage{bm}
\usepackage{rotating}
\usepackage[usenames, dvipsnames]{color}
\usepackage{hyperref}
\usepackage{cancel}
\usepackage{verbatim}
\usepackage{mathtools}
\usepackage{graphicx}
\usepackage{tensor}
\usepackage{verbatim}
\usepackage{tikz}
\usepackage{booktabs}
\graphicspath{ {images/} }

\usepackage{braket}

\newcommand{\ii}{\mathrm{i}}

\renewcommand{\d}{\mathrm{d}}
\renewcommand{\vec}{\text{vec}}
\newcommand{\op}[1]{\hat {#1}}

\begin{document}
\title{The Unruh effect in slow motion}

\author{Silas Vriend}
\email{vriendsp@mcmaster.ca}
\affiliation{Dept. of Math. \& Stat., McMaster University, Hamilton, ON, L8S 4S8, Canada}

\author{Daniel Grimmer}
\email{daniel.grimmer@philosophy.ox.ac.uk}
\affiliation{University of Oxford, Pembroke College, St. Aldates, Oxford, OX1 1DW, UK}

\author{Eduardo Mart\'{i}n-Mart\'{i}nez}
\email{emartinmartinez@uwaterloo.ca}
\affiliation{Institute for Quantum Computing, University of Waterloo, Waterloo, ON, N2L 3G1, Canada}
\affiliation{Dept. Applied Math., University of Waterloo, Waterloo, ON, N2L 3G1, Canada}
\affiliation{Perimeter Institute for Theoretical Physics, Waterloo, ON, N2L 2Y5, Canada}

\begin{abstract}
We show under what conditions an accelerated detector (e.g., an atom/ion/molecule) thermalizes while interacting with the vacuum state of a quantum field in a setup where the detector's acceleration alternates sign across multiple optical cavities. We show (non-perturbatively) in what regimes the probe `forgets' that it is traversing cavities and thermalizes to a temperature  proportional to its acceleration. Then we analyze in detail how this thermalization relates to the renowned Unruh effect. Finally, we use these results to propose an experimental testbed for the direct detection of the Unruh effect at relatively low probe speeds and accelerations, potentially orders of magnitude below previous proposals.
\end{abstract}

\maketitle

{\bf\textit{Introduction.-}}
The Unruh effect~\cite{Fulling1972,Davies1974,Unruh1976}, one of the fundamental and yet still untested predictions of quantum field theory, was first described as a consequence of the fact that the vacuum state of a quantum field in Minkowski space-time is  thermal (i.e., KMS) with respect to the generators of boosts~\cite{Takagi}. This translates to the fact that a uniformly accelerated observer of this vacuum will actually observe a thermal state whose temperature is proportional to her acceleration~\cite{Takagi,Crispino2007}. Later on, it was discussed that the Unruh effect is best described as the thermal response of an accelerated particle detector coupled to the vacuum~\cite{UnruhWald,Earman2011-EARTUE,Carballo2019}. This more modern view on the Unruh effect (taking the philosophy of ``particles are what particle detectors detect'') was first discussed by Unruh and Wald in~\cite{UnruhWald,Earman2011-EARTUE}.

Direct detection of the Unruh effect would be a feat that resonates across many fields, ranging from astrophysics \cite{Astronature,Hawking}, cosmology \cite{Cosmo,Cosmo2}, black-hole physics~\cite{Bholes}, particle physics~\cite{Base}, and quantum gravity~\cite{Qg,Hossain_2016,Rovelli2014} to the very foundations of QFT. Unsurprisingly, much effort has been made towards finding evidence of the Unruh (and the closely related Hawking) effect, both through direct and indirect observations~\cite{MartMart2011,PhysRevLett.83.256,PhysRevLett.118.161102} as well as in analog systems such as fluids~\cite{Unruhan}, Bose-Einstein condensates~\cite{garay,Boseando,Steinhauer_2016}, optical fibers \cite{optfib}, slow light \cite{slowlight},  superconducting circuits \cite{supercond} and trapped ions \cite{Milburn,ions}, to name a few.  Despite its fundamental relevance, an uncontroversial direct confirmation of the \mbox{Unruh} effect remains elusive.

In recent times, it has been shown that the Unruh effect may manifest itself even when the field state is not KMS with respect to accelerated observers~\cite{Carballo2019}. Roughly speaking, this is related to the fact that the only physical Lorentz invariant state of a free field in flat-spacetime is the vacuum, and that any deviations from the vacuum would eventually be red/blue-shifted out of the response window of any physical detector. The Unruh effect understood in terms of thermalization of particle detectors is a robust phenomenon. Indeed, one can even see this effect in settings (like optical cavities) where Lorentz invariance is explicitly broken~\cite{BrennaTherm}.

One commonality of all presently known scenarios exhibiting the Unruh effect is that the probe system becomes ultrarelativistic. This may seem unavoidable since the probe must accelerate for a long time (long enough to thermalize). Indeed, a na\"ive scale analysis suggests this. If a probe accelerates at a rate, $a$, and requires a proper time, $\tau_\text{thermal}$, to thermalize then it will achieve a Lorentz factor of \mbox{$\gamma_\text{thermal}=\text{cosh}(a \tau_\text{thermal}/c)$} with respect to its initial rest frame. If $a \tau_\text{thermal}/c\gtrsim10$ then $\gamma_\text{thermal}\gg1$.

However, as we will show in this manuscript, ultrarelativistic velocities from the probe's initial rest frame are not necessary to see the Unruh effect. We will show that we can achieve the thermalization of an accelerated particle detector to a temperature proportional to its acceleration as it interacts with the vacuum state of a field in a series of Dirichlet cavities at speeds as low as $v=0.6\,c\leftrightarrow\gamma=5/4$. Furthermore, our setup can in principle be used for an experimental detection of the Unruh effect requiring accelerations orders of magnitude smaller than the best current proposals known to the authors.

{\bf\textit{Motivation.-}} In the above argument that ultrarelativistic motion is unavoidable for the Unruh effect there is a hidden assumption: that the thermalization time is the same as the time that the acceleration is sustained. We can get around this by separating the two timescales. For instance, we can take the probe to alternate the sign of its acceleration at some regular interval, $\tau_\text{max}\ll\tau_\text{thermal}$. In this way the probe maintains a constant magnitude of acceleration, $\vert a\vert$, but does not accumulate much speed.

Will the probe still thermalize to the Unruh temperature when following this alternately accelerated/decelerated trajectory? One may have the intuition that it will since the probe would ``see a thermal bath of temperature $T_\textsc{u}= \hbar \vert a\vert/2\pi c k_\textsc{b}$'' between each acceleration sign-change event. If the probe does not thermalize it must be due to the sudden jerks felt by the probe at each acceleration sign-change event (or due to radiation produced at these events). 

This issue can be addressed by the following alternative setting: we set up a series of adjacent Dirichlet cavities containing quantum fields in their respective vacua. The walls of each cavity have small (say atom-sized) holes that the probe travels through. We take the probe to switch the sign of its acceleration as it crosses each cavity wall.

The benefits of introducing these cavity walls are three-fold. Firstly, the cavity walls reduce the continuum of field modes to a discrete collection of cavity modes, easier for computational purposes. 

Secondly, since the probe's interaction with the field is identical in each two-cavity-cell, we need only simulate the field-probe interaction for a relatively short duration, \mbox{$\delta t=2 \tau_\text{max}\ll\tau_\text{thermal}$}. Indeed, the cavity walls shield the probe from any radiation produced in previous cavities. As we will discuss in detail later, this makes the probe's dynamics time-independent and Markovian which allows for efficient non-perturbative calculations.

Thirdly, the field's boundary conditions enforce that the field amplitude vanishes at the cavity walls such that the probe is effectively decoupled from the field at each acceleration sign-change event. This eliminates the sudden jerks' effects on the probe's trajectory.

One may be concerned that these cavity walls will spoil the Unruh effect, for two main reasons. First, the vacuum in the cavity is not Lorentz invariant: there is a discrete set of field modes and the probe can notice this difference. Second, the probe creates disturbances in the field that will bounce off the cavity walls and affect the probe in turn.

We will see below that while there are indeed regimes where these effects prevent the probe from seeing the Unruh effect, there are also regimes where the probe is blind to the fact that it is in a cavity and experiences thermalization according to Unruh's law.

{\bf\textit{Our Setup.-}}
Consider a probe which is initially comoving with the cavity wall at $x=0$ and then begins to accelerate at a constant rate $a>0$ towards the far end of the cavity at $x=L>0$. In terms of the probe's proper time, $\tau$, this portion of the trajectory is given by
\begin{align}
\label{xtraj}
x(\tau)&=\frac{c^2}{a}(\text{cosh}(a \tau/c)-1),
\quad
t(\tau)=\frac{c}{a}\text{sinh}(a \tau/c), 
\end{align}
for \mbox{$0\leq \tau < \tau_\text{max}=\frac{c}{a}\text{cosh}^{-1}(1+a L/c^2)$}. The cavity-crossing time in the lab frame is  \mbox{$t_\text{max}=\frac{L}{c}\sqrt{1+2c^2/aL}$}. The probe exits the first cavity at some speed, $v_\text{max}$, relative to the the cavity walls with maximum Lorentz factor \mbox{$\gamma_\text{max}=\text{cosh}(a\tau_\text{max}/c)=1+aL/c^2$}. 

At $\tau=\tau_\text{max}$ the probe enters the second cavity of the two-cavity cell and begins decelerating with proper acceleration $a$. The probe reaches the far end of the second cavity, $x=2L$, just as it comes to rest at \mbox{$\tau=2\tau_\text{max}$}.

While a full light-matter interaction description would require a $3+1$D setup~\cite{Lopp_2018}, as proof of principle we will assume that each cavity contains a $1+1$D massless scalar field, $\op\phi(t,x)$, with a free Hamiltonian
\begin{align}
\op H_\phi = \frac{1}{2} \int_{0}^{L} \d x \: c^2 \op\pi(t,x)^2 + (\partial_x \op\phi(t,x))^2,
\end{align}
satisfying \mbox{$[\op\phi(t,x), \op\pi(t,x')] = \ii\hbar \delta(x - x') \op\openone$},  where $ \op{\pi}(t,x) $ is the field's canonical conjugate momentum. The field obeys Dirichlet boundary conditions at $x=0$ and $x=L$ such that,
\begin{align}
\op{\phi}(t,x) \!=\! \sum_{n = 1}^\infty &\sqrt{\frac{ 2 \hbar c^2}{\omega_n L}}\sin(k_n x)
\left(\op{a}^\dagger_n e^{\ii\omega_n t} + \op{a}_n e^{-\ii\omega_n t}\right)\!,
\end{align}
where mode frequencies and wavenumbers satisfy \mbox{$ck_n = \omega_n = n c \pi/L$}, and $\op{a}^\dagger_n,\,\op{a}_n$ are  the $ n^\mathrm{th}$-mode's creation/annihilation operators. 

Let the probe's internal degree of freedom be a quantum harmonic oscillator with some energy gap, $\hbar\Omega_\textsc{p}$. The probe is characterized by dimensionless quadrature operators $\hat{q}_\textsc{p}$ and $\hat{p}_\textsc{p}$ obeying \mbox{$[\op q_\textsc{p}, \op p_\textsc{p}] = \ii \op\openone$}. In these terms the probe's free Hamiltonian is \mbox{$\hat{H}_\textsc{p}=\hbar\Omega_\textsc{p} (\hat q_\textsc{p}^2+\hat p_\textsc{p}^2-1)/2$}. In the interaction picture $\op{q}_\textsc{p}(\tau)$ evolves with respect to $\tau$ as \mbox{$\op{q}_\textsc{p}(\tau) = \op{q}_\textsc{p}(0) \cos(\Omega_\textsc{p} \tau) + \op{p}_\textsc{p}(0) \sin(\Omega_\textsc{p} \tau)$}.

We take the probe to couple to the field via the Unruh-DeWitt interaction Hamiltonian~\cite{DeWitt1980,Takagi,Crispino2007}, 
\begin{align}
\op{H}_I(\tau) = \lambda \ \op q_\textsc{p}(\tau) \ 
\op\phi\!\left(t(\tau), x(\tau)\right),
\end{align}
where $\lambda$ is the coupling strength. This Hamiltonian captures the fundamental features of the light-matter interaction when exchange of angular momentum is not relevant~\cite{cohen,scully_zubairy_1997,pablooptics,lopp2020quantum}. Note that $x(\tau)$ and $t(\tau)$ are given by Eq. \eqref{xtraj} while the probe accelerates through the first cavity. The trajectory in the second cavity of the cell is a straightforward reversed-translation of this trajectory.

{\bf\textit{Non-perturbative time-evolution.-}} We next compute the probe's dynamics in the first cell. In the interaction picture the time-evolution operator for the probe-field system in the $n^\text{th}$ cavity is,
\begin{align}
\op U_n^\text{I}= \mathcal{T}\!\exp \left(\frac{-\ii}{\hbar}\int_{(n-1)\tau_\text{max}}^{n\tau_\text{max}}\d \tau \op H_I(\tau)\right).
\end{align}
The probe's reduced dynamics is given by,
\begin{align}
    \Phi_n^\text{I}[\op\rho_\textsc{p}] = \text{Tr}_{\phi}(\op U_n ^\text{I} (\op\rho_\textsc{p} \otimes \ket{0}\!\bra{0}) \op U_n ^\text{I}{}^\dagger).
\end{align}
Composing the cases $n=1$ and $n=2$ (where the probe accelerates and decelerates respectively) we can build the interaction picture update map for the first cell, \mbox{$\Phi ^\text{I}_{1,2}=\Phi_2 ^\text{I} \circ \Phi ^\text{I}_1$}.

Analogously, one can find the update map for the second cell, $\Phi ^\text{I}_{3,4}=\Phi_4 ^\text{I} \circ \Phi_3 ^\text{I}$, but unfortunately this map is different for every cell (\mbox{$\Phi ^\text{I}_{3,4}\neq\Phi ^\text{I}_{1,2}$}).
However in the Schr\"odinger picture the update map is in fact the same for each cell, \mbox{$\Phi ^\text{S}_\text{cell}=\Phi ^\text{S}_{1,2}=\Phi ^\text{S}_{3,4}=\dots$}. We can build $\Phi ^\text{S}_\text{cell}$ from the above discussed update maps as \mbox{$\Phi ^\text{S}_\text{cell}=\mathcal{U}_0^2 \circ \Phi_2 ^\text{I} \circ \Phi_1 ^\text{I}$} where \mbox{$\mathcal{U}_0[\hat\rho_\textsc{p}]=U_0 \hat\rho_\textsc{p} U_0^\dagger$} and \mbox{$U_0=\exp(-\ii  \tau_\text{max} \hat{H}_\textsc{p}/\hbar)$} (see auxiliary technical details in Appendix \ref{AppSec1}).

In summary, as the probe travels through many cells it is repeatedly updated by $\Phi ^\text{S}_\text{cell}$. Noting that $\Phi ^\text{S}_\text{cell}$ depends on the cell-crossing time, $\delta t=2 \tau_\text{max}$, we have, 
\begin{align}\label{UpdateEquation}
\hat\rho_\textsc{p}(n\,\delta t)=\big(\Phi_\text{cell} ^\text{S}(\delta t)\big)^n[\hat\rho_\textsc{p}(0)].
\end{align}
This dynamics is Markovian and time-independent: the same update map is applied each time-step.

There are powerful tools to analyze the dynamics of such repeated update systems. One such tool is the \textit{Interpolated Collision Model} formalism, ICM~\cite{GrimmerThesis,RRI,Pure}, which allows us to rewrite the discrete update equation~\eqref{UpdateEquation}~as a differential equation with no approximation and without needing to take $\delta t\to0$ unlike in other common approaches~\cite{Giovannetti:2012, CM,ACMZ,PhysRevLett.115.120403,PhysRevLett.108.040401,PhysRevA.98.062104,PhysRevA.97.053811,PhysRevX.7.021003,PhysRevA.96.032107,PhysRevA.95.053838,Giovannetti_2012,Altamirano_2017,Attal2007,doi:10.1063/1.4879240}.

Additionally, we take advantage of the fact that our setup is Gaussian: all the states involved have Gaussian Wigner functions and interact through quadratic Hamiltonians. This enables us to simplify our description of the probe's state from an infinite dimensional density matrix, $\hat\rho_\textsc{p}(n \delta t)$, to just a $2\times2$ covariance matrix, $\sigma_\textsc{p}(n \delta t)$, for the probe's quadrature operators; see~\cite{AdessoThesis,GQMRev,lami,GaussianClass}.

Using recent results on Gaussian ICM~\cite{GrimmerThesis,GAB} we can efficiently calculate the fixed points and convergence rates of repeated application of $\Phi ^\text{S}_\text{cell}$. This is achieved by straightforward application of the formalism developed in~\cite{GAB}. For the convenience of the reader we provide a quick summary particularized to our setup in Appendix~\ref{AppSec2}.

{\bf\textit{Results.-}}
As we have discussed above, we can efficiently compute the probe's final covariance matrix, $\sigma_\textsc{p}(\infty)$, after it has traveled through many cells. $\sigma_\textsc{p}(\infty)$ is the unique fixed point of $\Phi ^\text{S}_\text{cell}$. To characterize this state we write it in standard form,
\begin{align}\label{FinalState}
\sigma_\textsc{p}(\infty)
=\mathcal{R}(\theta)\begin{pmatrix}
\nu \exp(r) & 0 \\
0 & \nu \exp(-r) \\
\end{pmatrix}\mathcal{R}(\theta)^\intercal,
\end{align}
for some symplectic eigenvalue $\nu\geq1$, squeezing parameter $r\geq0$ and angle $\theta\in[-\pi/2,\pi/2]$ where $\mathcal{R}(\theta)$ is the $2\times2$ rotation matrix. The questions that we will answer next are: a) is the probe's final state thermal? and if so, b) how does the probe's final temperature depend on the parameters of our setup? 

The free parameters are: 1)--the cavity length, $L$, 2)--the probe's proper acceleration, $a$, 3)--the probe's proper frequency $\Omega_\textsc{p}$, and 4)--the coupling strength, $\lambda$. The relevant dimensionless variables are \mbox{$a_0=a L/c^2,\,\Omega_0=\Omega_\textsc{p} L/c,\, \text{ and }\lambda_0=\lambda L/\sqrt{\hbar c}$}. We fix \mbox{$\lambda_0=0.01$}, but our results are independent of the coupling strength provided $\lambda_0\lesssim1$.

We next investigate for what values of $a_0$ and $\Omega_0$ the final probe state is approximately thermal. From~\eqref{FinalState}, if the probe state is not squeezed (i.e., $r=0$) then it is in a thermal state with temperature $
k_\textsc{b} T = \hbar\Omega_\textsc{p}/2 \text{arccoth}(\nu)$. It is intuitive that if $r$ is ``small enough'' then we can say the state is approximately thermal. The question is then ``how small is small enough?'' One possible estimate of thermality is to compare how the energy to build the state from the vacuum splits between energy spent on squeezing versus heating (see e.g.~\cite{BrennaTherm}). For the regimes where we see the Unruh effect, this ratio is always less than $0.001\%$. For the interested reader, we consider several different temperature estimates and measures of thermality in Appendix \ref{AppSec3}. Over the parameter range considered in this manuscript these measures of thermality all indicate that the probe's final state is effectively indistinguishable from thermal. 

\begin{figure}[t]
	\centering
	\includegraphics[width=0.95\linewidth]{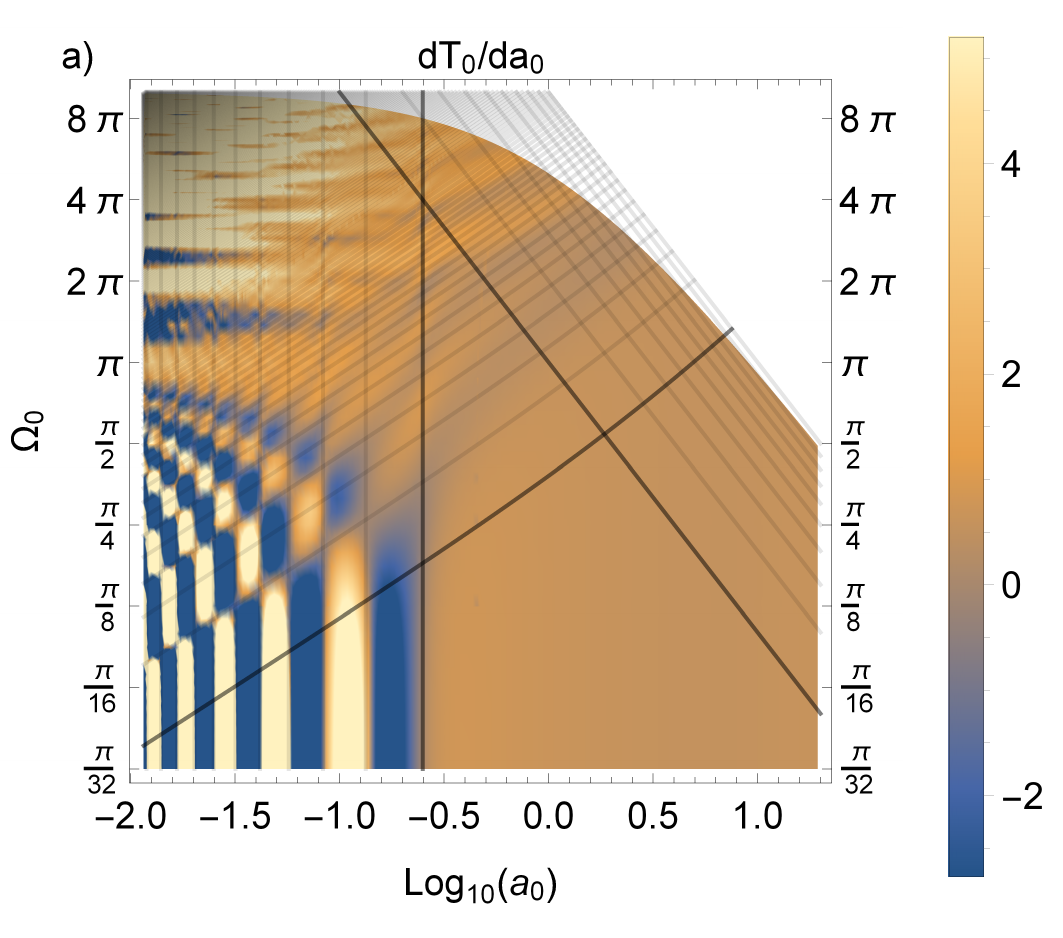}
	\includegraphics[width=0.95\linewidth]{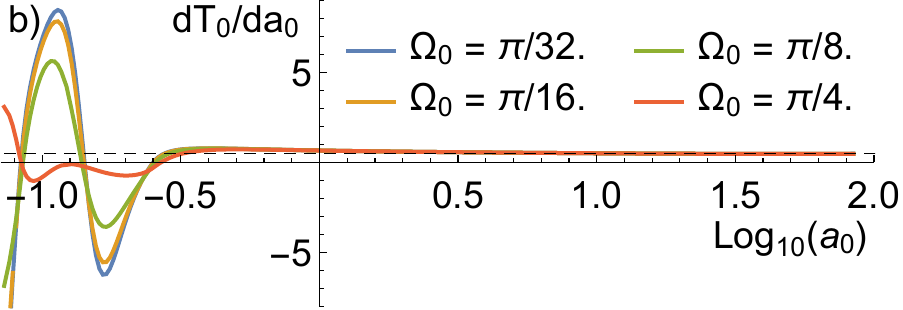}
	\caption{a) Derivative of the  probe's final temperature \mbox{$T_0=k_\textsc{b} T   L/\hbar c$} with respect to the  acceleration \mbox{$a_0= a L/c^2$} as a function of $a_0$ and the probe gap \mbox{$\Omega_0=\Omega_\textsc{p} L/c$} on log-scale. The dimensionless coupling strength is fixed at $\lambda_0=\lambda L/\sqrt{\hbar c}=0.01$. The Unruh effect ($\d T/\d a\approx\text{constant}$) is found in the lower-right portion of the figure. Black lines highlight relevant physical scales (see main text). b) Cross-sections of $\d T_0/\d a_0$ as a function of $a_0$  for several detector gaps: $\Omega_0=\pi/32, \ \pi/16, \ \pi/8, \ \pi/4$ (from top to bottom at $\text{Log}_{10}(a_0)=-1$)) showing independence of $\Omega_0$ in the Unruh effect regime. The black-dashed line is at $\d T_0/\d a_0=1/2$.}
	\label{dTdaFig}
\end{figure}

Since the probe is indistinguishable from thermal, we next ask how its (dimensionless) final temperature, \mbox{$T_0 = k_\textsc{b} T   L/\hbar c$}, depends on $a_0$ and $\Omega_0$. A clear signature of the Unruh effect would be finding $T\propto a$. We thus search for regimes where $\d T_0/\d a_0$ is constant (i.e., independent of both $a_0$ and $\Omega_0$). Fig.~\ref{dTdaFig}a shows $\d T_0/\d a_0$ for a wide range of accelerations and probe gaps. Note that we approach a constant value of $\d T_0/\d a_0$ in the bottom-right of the figure. 

The upward sloping lines in Fig.~\ref{dTdaFig}a indicate the parameters for which the probe's free Hamiltonian rotates through a phase of  $\Theta=\Omega_\textsc{p} \tau_\text{max}=n \pi/2$ inside each cavity. The $\Theta=\pi/2$ line is bold. There are fundamental limits to the energy resolution that detectors can achieve coming from energy-time uncertainty principles~\cite{Smith2020}. To resolve the cavity into discrete energy levels any detector would need to interact for a time long enough to allow its internal energy uncertainty to decrease to a point where it can confidently distinguish between two different discrete levels.  For our detector this means $\Theta\gg 2\pi$. Note that the regime where $\d T_0/\d a_0\approx\text{constant}$ is located below the $\Theta=\pi/2$ line such that in this regime the probe cannot fully resolve the cavity into discrete levels. 

Resolving the cavity's discrete spectrum is not the only way that the probe could learn that it is in a cavity. Indeed, the probe may learn of the cavity walls by bouncing a signal off of them. Consider the disturbances that the detector is effecting on the field as it goes along its trajectory. If initially right-moving (left-moving), these disturbances cross paths with the probe an odd (even) number of times. In each case the minimum number of crossings is achieved for $M\leq3$ where $M=c t_\text{max}/L$ is the ratio of the probe's cavity-crossing time, $t_\text{max}$, to the cavity's light-crossing time, $L/c$. The vertical lines in Fig.~\ref{dTdaFig}a correspond to $M=3$ (bold) and $M=4,\,5,\,6,\,\dots$. Note that the regime where $\d T_0/\d a_0\approx\text{constant}$ is located to the right of the $M=3$ line (i.e., for $a_0>1/4$). In this region the probe does not spend long in each cavity (less than three light-crossing times) and thereby interacts minimally with any reflected signals.

Interestingly, the probe's gap, $\Omega_\textsc{p}$, does not need to sweep across many cavity modes as it is blue/red-shifted ($\Omega_\textsc{p}\leftrightarrow\gamma_\text{max} \Omega_\textsc{p}$) with respect to the lab frame. The number of cavity modes in this range is approximately \mbox{$R=(\gamma_\text{max}\Omega_\textsc{p}-\Omega_\textsc{p})/(\omega_2-\omega_1)$}. The downward sloping lines in Fig.~\ref{dTdaFig} are located at \mbox{$R=1,2,3\dots$} with $R=1$ bold. Note that the regime where \mbox{$\d T_0/\d a_0\approx\text{constant}$} is on either side of the $R=1$ line. Note, however, that all our calculations always consider a large enough number of modes $N\gg R$ in the cavity to guarantee that the error in the sum is negligible (See Appendix \ref{ModeConvergence} for details on mode-convergence).

Summarizing,  Fig~\ref{dTdaFig}a and Fig~\ref{dTdaFig}b show that above \mbox{$a_0=1/4$} and below \mbox{$\Theta=\pi/2$} we have $\d T_0/\d a_0\approx1/2$. The detector thermalizes to a temperature which is proportional to its acceleration and independent of $\Omega_0$, the hallmark of the Unruh effect. The only difference with the continuum Unruh effect is that there \mbox{$\d T_0/\d a_0 = 1/2\pi$}. 

{\bf\textit{The missing pie.-}}
Undoubtedly this mismatch of slopes ($1/2 \text{ vs } 1/2\pi$) is a glaring difference between this Unruh effect in many cavities and the canonical one in the continuum. We account for this difference by noting that there is no limit in which our setup returns the canonical Unruh effect scenario. It is critical in our setup that the probe does not have time to resolve the cavity into discrete energy levels, i.e., that $\Theta=\Omega_\textsc{p}\tau_\text{max}\lesssim 2\pi$. This precludes the probe from thermalizing within a single cavity, since \mbox{$\tau_\text{max}\lesssim 2\pi/\Omega_\textsc{p}$} is less than the probe's Heisenberg time. Thus, in our setup the probe's thermalization is necessarily a multi-cavity phenomenon, making it unachievable in the $L\to\infty$ limit and hence difficult to compare with the continuum.

The exact magnitude of the slope may be capturing geometric factors (that are dimension dependent, yielding missing $\pi$'s) and/or the scales we have fixed e.g., the probe's initial velocity. However, we will still argue along the lines of~\cite{UnruhWald,Carballo2019} that the most fundamental part of the Unruh effect is that an accelerated detector interacting with the ground state of a quantum field thermalizes in the long-time limit to a temperature proportional to its acceleration regardless of its internal energy-gap and the coupling strength. 

{\bf\textit{Experimental detection.-}}  Our proposal can achieve the Unruh effect for dimensionless accelerations as small as $a_0=a L/c^2=1/4$ where $L$ is the cavity length. For the probe, this corresponds to a maximum Lorentz factor of $\gamma_\text{max}=1+a_0=5/4$ with respect to the cavity walls (a speed of $v=0.6 c$).

Proposing a particular implementation of an Unruh effect experiment is outside the scope of this manuscript, but this setting can provide, at least in principle, a much better testbed for future experimental explorations. For instance, for a table-top setup with $L=1 \text{ m}$ this is an acceleration of $a=2.3\times 10^{15} g$. This matches the lowest-acceleration experimental proposals for direct detection known to the authors~\cite{MartMart2011,PhysRevLett.83.256,PhysRevLett.118.161102}. At this scale the resonance with the first cavity mode \mbox{($\Omega_0=\pi$)} happens at probe frequency \mbox{$\Omega_\textsc{p}= 1\text{ GHz}$}, (e.g., the Hydrogen 21-cm line).

For the largest cavity on Earth (LIGO, $L=4\text{ km}$) we can lower the required acceleration way below any previous proposal to $a=5.7\times 10^{11} g$. At this scale $\Omega_0=\pi$ happens at a frequency of $\Omega_\textsc{p}=240\text{ kHz}$, well within molecular transitions in the sub-Doppler regime (e.g.,~\cite{foltynowicz2020subdoppler}).

{\bf\textit{Conclusions.-}}
We have presented a setup which displays the Unruh effect (thermalization of a particle detector to a temperature proportional to its acceleration) without the detector becoming ultrarelativistic. Moreover, this setup has the potential to provide an experimental testbed for the Unruh effect orders of magnitude lower than previous proposals. 

We achieved this by having the probe alternate between accelerating and decelerating at regular intervals. To make this non-perturbative calculation feasible we consider the probe moving within a series of optical cavities. The alternation of acceleration and deceleration in different cavities makes the probe's dynamics the result of a repeated update map, allowing us to use the non-perturbative ICM formalism without any approximations or assumptions~\cite{GrimmerThesis}. 

Despite the departures from the canonical Unruh effect scenario (the vacuum of a free field in a cavity is not Lorentz invariant) we still see the Unruh effect (as in~\cite{BrennaTherm}) and further discuss that when the Unruh effect is present it is because the probe does not have enough time to learn that it is in cavity (either by resolving the cavity's discrete energy levels or by bouncing a signal off the walls). In this regime, the probe thermalizes to an Unruh temperature with the cavities collectively despite not having time to thermalize with each one individually.



{\bf\textit{Acknowledgements.-}}
The authors thank Jose De Ramon (Pipo) for illuminating discussions.  EMM acknowledges support through the Discovery Grant Program of the Natural Sciences and Engineering Research Council of Canada (NSERC). EMM also acknowledges support of his Ontario Early Researcher award. DG acknowledges support by NSERC through a Vanier Scholarship. SV acknowledges support by NSERC through a CGS M award. This work was made possible by the facilities of the Shared Hierarchical Academic Research Computing Network (SHARCNET:www.sharcnet.ca) and Compute/Calcul Canada.

\newpage

\onecolumngrid

\appendix

\section{Single-Cell Dynamics in the Interaction and Schr\"odinger Pictures}\label{AppSec1}

As we discussed in the main text the update map for the probe crossing one cell is best viewed in the Schr\"odinger picture whereas the dynamics is easiest to compute in the interaction picture. In this section we will lay out the details of how these pictures relate to each other for our setup. 

In the Schr\"odinger picture the time evolution operator from the start of the $n^\text{th}$ cavity (at $\tau=(n-1)\tau_\text{max}$) to the end of the $n^\text{th}$ cavity (at $\tau=n\tau_\text{max}$) is given by,
\begin{align}
    \op U^\text{S}_n = \mathcal{T}\!\exp \left(\frac{-\ii}{\hbar}\int_{(n-1)\tau_\text{max}}^{n\tau_\text{max}}\d \tau \ \op H_0(\tau)+H_I^\text{S}(\tau)\right),
\end{align}
where $\hat{H}_0(\tau)=\hat{H}_\textsc{p}+\frac{\d t}{\d \tau}\hat{H}_\phi$ is the sum of the probe and field's free Hamiltonians and $\hat{H}_I^\text{S}(\tau)=\lambda\,\hat{q}_\textsc{p}\otimes\hat\phi(x(\tau))$ is the probe-field interaction Hamiltonian in the Schr\"odinger picture. Note that since the field's free Hamiltonian generates evolution with respect to the lab time, $t$, it is modified by the time dilation factor $\d t/\d \tau$ in the above expression~\cite{pablooptics}.

We note that the above unitary only depends on whether $n$ is even or odd; that is, whether the probe is accelerating or decelerating. For example, the probe-field interaction in the third cavity is identical to the interaction in the first cavity, just shifted in space and time. Thus we only need to calculate,
\begin{align}\label{AppEq1}
\op U^\text{S}_+\coloneqq\op U^\text{S}_1=\op U^\text{S}_3=\op U^\text{S}_5=\dots
\quad\text{and}\quad
\op U^\text{S}_-\coloneqq\op U^\text{S}_2= \op U^\text{S}_4= \op U^\text{S}_6=\dots,
\end{align}
to fully specify the dynamics.
The subindices $+$ and $-$ correspond to cavities where the probe is accelerating and decelerating, respectively. Once we have computed $\op U^\text{S}_+$ and $\op U^\text{S}_-$ we can then compute the reduced maps for the probe in the Schr\"odinger picture as,
\begin{align}
\Phi_+^\text{S}[\op\rho_\textsc{p}] = \text{Tr}_{\phi}(\op U^\text{S}_+ (\op\rho_\textsc{p} \otimes \ket{0}\!\bra{0}) \op U^\text{S}_+{}^\dagger),
\quad\text{and}\quad
\Phi_-^\text{S}[\op\rho_\textsc{p}] = \text{Tr}_{\phi}(\op U^\text{S}_- (\op\rho_\textsc{p} \otimes \ket{0}\!\bra{0}) \op U^\text{S}_-{}^\dagger).
\end{align}
The update map for every cell is then $\Phi^\text{S}_\text{cell}=\Phi^\text{S}_-\circ\Phi^\text{S}_+$ in the Schr\"odinger picture. As such, the probe's state when it exits the $n^\text{th}$ cell (at proper time $\tau=n\delta t$ where $\delta t=2\,\tau_\text{max}$) is given by,
\begin{align}\label{AppEq3}
\hat\rho^\text{S}_\textsc{p}(n\,\delta t)=
(\Phi^\text{S}_\text{cell})^n[\op\rho_\textsc{p}(0)],
\end{align}
as claimed in the main text.

While the above update map is straightforwardly defined it is not the easiest to compute. It is much easier to compute the analogous unitaries in the interaction picture,
\begin{align}\label{AppEq2}
    \op U^\text{I}_n = \mathcal{T}\!\exp \left(\frac{-\ii}{\hbar}\int_{(n-1)\tau_\text{max}}^{n\tau_\text{max}}\d \tau \op H_I^\text{I}(\tau)\right),
\end{align}
where $\hat{H}_I^\text{I}(\tau)=\lambda \ \op q_\textsc{p}(\tau) \, \otimes \,
\op\phi\!\left(t(\tau), x(\tau)\right)$ is the probe-field interaction Hamiltonian in the interaction picture. From this we can construct the update map for the $n^\text{th}$ cavity in the interaction picture,
\begin{align}\label{AppEq4}
\Phi_n^\text{I}[\op\rho_\textsc{p}] = \text{Tr}_{\phi}(\op U^\text{I}_n (\op\rho_\textsc{p} \otimes \ket{0}\!\bra{0}) \op U^\text{I}_n{}^\dagger).
\end{align}
We can then convert these to the Schr\"odinger picture using the free evolution operator. The free evolution unitary operator for the $n^\text{th}$ cavity is,
\begin{align}
    \op V_{0,n} &= \mathcal{T}\!\exp \left(\frac{-\ii}{\hbar}\int_{(n-1)\tau_\text{max}}^{n\tau_\text{max}}\d \tau \op H_0(\tau)\right)\\
    &= \mathcal{T}\!\exp \left(\frac{-\ii}{\hbar}\int_{(n-1)\tau_\text{max}}^{n\tau_\text{max}}\d \tau \op H_\textsc{p}\right)
    \otimes
    \mathcal{T}\!\exp \left(\frac{-\ii}{\hbar}\int_{(n-1)t_\text{max}}^{n\,t_\text{max}}\d t \, \op H_\phi\right)\\
    &= \exp \left(-\ii\,\tau_\text{max}\, \op H_\textsc{p}/\hbar\right)
    \otimes
    \exp \left(-\ii\,t_\text{max}\, \op H_\phi/\hbar\right)\\
    &= \hat{U}_0 \otimes \hat{W}_0,
\end{align}
where $\hat{U}_0=\exp(-\ii\,\tau_\text{max}\, \op H_\textsc{p}/\hbar)$ and $\hat{W}_0=\exp(-\ii\,t_\text{max}\, \op H_\phi/\hbar)$. Thus the free evolution operator for each cavity is independent of $n$ and is a tensor product, so we may write $\hat{V}_0\coloneqq \hat{U}_0 \otimes \hat{W}_0$. For later convenience we will also define the maps $\mathcal{V}_0[\hat{\rho}]= \hat{V}_0 \ \hat{\rho} \ \hat{V}_0^\dagger$ and $\mathcal{U}_0[\hat{\rho}_\textsc{p}]= \hat{U}_0 \ \hat{\rho}_\textsc{p} \ \hat{U}_0^\dagger$.

Now that we have computed the free evolution operator we can use it to write the interaction picture unitaries, $\op U^\text{I}_n$, in terms of their Schr\"odinger picture counterparts, $\op U^\text{S}_n$, as,
\begin{align}\label{Equation}
\hat{U}^\text{I}_n&=(\hat{V}_0^\dagger)^n \, \hat{U}^\text{S}_n \, (\hat{V}_0)^{n-1}.
\end{align}
Note that $\hat{U}^\text{I}_n$ depends on $n$ in two ways, through $\hat{U}^\text{S}_n$ and through the number of free rotations, $\hat{V}_0$, to be applied. The first kind of dependence is the same as in the Schr\"odinger picture case (i.e., dependence on whether the probe is accelerating or decelerating through the $n^\text{th}$ cavity). The second kind of dependence is new: it is due to the time-dependence brought about by $\hat V_0$ in the interaction picture. The dictionary between the Schr\"odinger and interaction pictures is itself time-dependent. This dependence can be seen in \eqref{AppEq2} by noting that the probe's quadrature operators are different at the beginning of each interaction, 
\begin{align}
\op q_\textsc{p}^\text{I}(0)
\neq\op q_\textsc{p}^\text{I}(\tau_\text{max})
\neq\op q_\textsc{p}^\text{I}(2\,\tau_\text{max})
\neq\dots 
\neq\op q_\textsc{p}^\text{I}(N\,\tau_\text{max}).
\end{align}
This second kind of dependence on $n$ ultimately prevents us from writing an update map of the form \eqref{AppEq3} in the interaction picture since the update map for each cell will be different. Thus if we would like to make use of the ICM formalism discussed in the main text, we need to work in the Schr\"odinger picture.

This does not mean however that computations done in the interaction picture are useless. Indeed we can construct the Schr\"odinger picture update map from $\Phi^\text{I}_1$ and $\Phi^\text{I}_2$ and $\mathcal{U}_0$ as follows. We first note that $\hat{U}^\text{S}_+$ and $\hat{U}^\text{S}_-$ can be written in terms of $\hat{U}^\text{I}_1$, $\hat{U}^\text{I}_2$, and $\hat{V}_0$ as,
\begin{align}
\hat{U}^\text{S}_+ = \hat{U}^\text{S}_1 =\hat{V}_0 \ \hat{U}^\text{I}_1,
\quad\text{and}\quad
\hat{U}^\text{S}_- = \hat{U}^\text{S}_2 = \hat{V}_0^2 \ \hat{U}^\text{I}_2 \ \hat{V}_0^\dagger,
\end{align}
where we have used \eqref{Equation} with $n=1$ and $n=2$, respectively. Recalling that $\hat{V}_0=\hat{U}_0\otimes \hat{W}_0$ and noting that the field's initial state, $\ket{0}\!\bra{0}$, is fixed under its free dynamics we then have,
\begin{align}
\Phi_+^\text{S}[\op\rho_\textsc{p}] &=  (\mathcal{U}_0\circ \Phi^\text{I}_1)[\op\rho_\textsc{p}],\\
\Phi_-^\text{S}[\op\rho_\textsc{p}] &=  (\mathcal{U}_0^2\circ \Phi^\text{I}_2 \circ \mathcal{U}_0^\dagger)[\op\rho_\textsc{p}].
\end{align}
Composing these two maps we find $\Phi_\text{cell}^\text{S}[\op\rho_\textsc{p}] 
= (\Phi^\text{S}_- \circ \Phi^\text{S}_+)[\op\rho_\textsc{p}]
= (\mathcal{U}_0^2\circ \Phi^\text{I}_2 \circ \Phi^\text{I}_1)[\op\rho_\textsc{p}]$ as claimed in the main text.

\section{Gaussian Interpolated Collision Model Formalism}\label{AppSec2}
As discussed in the main text, our ability to efficiently calculate the fixed points and convergence rates of repeated application of $\Phi ^\text{S}_\text{cell}$ is aided by two facts: our setup is both Gaussian and Markovian. This allows us to use Gaussian Quantum Mechanics (GQM) and more specifically the Gaussian Interpolated Collision Model formalism (Gaussian ICM) for our calculations. This section will briefly review those well-known techniques and show how they are applied to our setup. More details on GQM and Gaussian ICM can be found in \cite{AdessoThesis,GQMRev,lami,GaussianClass} and~\cite{GAB,GrimmerThesis},  respectively.

\subsection{Gaussian Quantum Mechanics}
GQM is a restriction of quantum mechanics in which we restrict ourselves to Gaussian states (states with Gaussian Wigner functions) and quadratic Hamiltonians. In GQM:
\begin{itemize}
    \item[1)] density matrices, $\hat\rho$, are replaced with covariance matrices, $\sigma$, and displacement vectors, $\bm{x}$, which fully characterize a Gaussian state in phase space;
    \item[2)] quadratic Hamiltonians, $\hat H$, are replaced with a quadratic form, $F$, and a vector, $\bm{\alpha}$, such that \mbox{$\hat H=\frac{1}{2}\hat{\bm{X}}^\intercal F\hat{\bm{X}}+\hat{\bm{X}}^\intercal \bm{\alpha}$}, where $\hat{\bm{X}}^\intercal=(\hat{q}_0,\hat{p}_0,\hat{q}_1,\hat{p}_1,\dots)$ is the vector of the system's quadrature operators;
    \item[3)] unitary evolution, $\hat\rho\to \hat U\,\hat\rho\, \hat U^\dagger$, is explicitly implemented as symplectic(-affine) evolution  $\sigma\to S\,\sigma\, S^\intercal$ and $\bm{x}\to S\bm{x}+\bm{d}$, where $S$ is a symplectic transformation; that is, $S$ is a transformation which preserves the symplectic form, $\Omega$, (defined via $[\hat{\bm{X}}_i,\hat{\bm{X}}_j]=\ii\Omega_{ij}\hat{\openone}$), in the sense that $S\Omega S^\intercal=\Omega$;
    \item[4)] as a consequence of the formalism, tensor products, $\hat\rho_\text{AB}=\hat\rho_\text{A}\otimes\hat\rho_\text{B}$, are replaced with (simpler) direct sums, $\sigma_\text{AB}=\sigma_\text{A}\oplus\sigma_\text{B}$. Correspondingly, partial traces are replaced with an analogous reduction map, $M$, such that $M_B(\sigma_\text{A}\oplus\sigma_\text{B})=\sigma_\text{A}$.
\end{itemize}
Concretely, the unitary transformation for the $n^\text{th}$ cavity in the interaction picture (Eq.~\eqref{AppEq2}),
\begin{align}
    \op U^\text{I}_n = \mathcal{T}\!\exp \left(\frac{-\ii}{\hbar}\int_{(n-1)\tau_\text{max}}^{n\tau_\text{max}}\d \tau \op H_I^\text{I}(\tau)\right),
\end{align}
gives rise to the symplectic transformation,
\begin{align}
    \op S^\text{I}_n = \mathcal{T}\!\exp \left(\frac{1}{\hbar}\int_{(n-1)\tau_\text{max}}^{n\tau_\text{max}}\d \tau \ \Omega F_I^\text{I}(\tau)\right),
\end{align}
where $\hat{H}_I^\text{I}(\tau)=\frac{1}{2}\hat{\bm{X}}^\intercal F_I^\text{I}(\tau)\hat{\bm{X}}$. This symplectic transformation is computationally more accessible than the corresponding unitary transformation. Recall that in the Hilbert space treatment each cavity mode corresponds to an infinite-dimensional factor in the full Hilbert space. Contrast this with the Gaussian treatment where each cavity mode corresponds to a two-dimensional subspace of the full phase space. Thus, if we can accurately simulate our setup using only a (possibly large but) finite number of cavity modes, $N$, then $F_I^\text{I}(\tau)$ is a finite-dimensional matrix (of dimension $2(N+1)$). If it were possible to address this scenario by considering enough cavity modes to have convergence this would make a non-perturbative calculation of the dynamics feasible. We will discuss the number of cavity modes needed for convergence in Sec. \ref{ModeConvergence}.

The update map for the $n^{th}$ cavity in the interaction picture (Eq.~\eqref{AppEq4}),
\begin{align}
\Phi_n^\text{I}:\quad\op\rho_\textsc{p}\to \text{Tr}_{\phi}(\op U^\text{I}_n (\op\rho_\textsc{p} \otimes \op\rho_\phi) \op U^\text{I}_n{}^\dagger),
\end{align}
can be understood to act on the probe's covariance matrix, $\sigma_\textsc{p}$, as,
\begin{align}\label{GaussianMap}
\Phi_n^\text{I}:\quad\sigma_\textsc{p} \to \text{M}_{\phi}(S^\text{I}_n (\sigma_\textsc{p} \oplus \sigma_\phi) S^\text{I}_n{}^\intercal).
\end{align}
That is, the probe's covariance matrix is embedded into a larger phase space, evolved symplectically, and finally projected back into its original phase space. Note that since the probe and field initially have no displacement, $\bm{X}_\textsc{p}(0)=0$ and $\bm{X}_\phi(0)=0$, and there are no linear terms in the Hamiltonian, $\alpha=0$, we have that $\bm{X}_\textsc{p}(t)=0$ and $\bm{X}_\phi(t)=0$ for all $t$. Thus we can restrict our attention to just the probe and field's covariance matrices.

It is worth noting that while $\Phi_n^\text{I}$ acts linearly on $\hat\rho_\textsc{p}$ it acts in a linear-affine way on $\sigma_\textsc{p}$. In fact, it is straightforward to rewrite \eqref{GaussianMap} in the form,
\begin{align}
\Phi^\text{I}_n:\quad\sigma_\textsc{p} &\to T_n^\text{I} \, \sigma_\textsc{p} \, T_n^\text{I}{}^\intercal +R_n^\text{I},
\end{align}
for some real $2\times2$ matrices  $T_n^\text{I}$ and $R_n^\text{I}$ which can be calculated directly from $S^\text{I}_n$ and $\sigma_\phi$.  As we discussed in the previous section, we only need to calculate $\Phi^\text{I}_n$ for $n=1$ and $n=2$ to fully specify the dynamics. That is, we only need to calculate $T_1^\text{I}$, $R_1^\text{I}$, $T_2^\text{I}$, and $R_2^\text{I}$ and then convert these to the Schr\"odinger picture in order to easily concatenate the different cell maps.

In order to convert these to the Schr\"odinger picture we need the Gaussian version of the probe's free evolution map, $\mathcal{U}_0$. This is given by,
\begin{align}\label{SuppU0}
\mathcal{U}_0:\quad\sigma_\textsc{p} &\to \mathcal{R}(\Omega_\textsc{p}\,\tau_\text{max}) \ \sigma_\textsc{p} \ \mathcal{R}(\Omega_\textsc{p}\,\tau_\text{max})^\intercal,
\end{align}
where $\mathcal{R}(\theta)$ is the $2\times 2$ rotation matrix. That is, in phase space, the probe's free evolution is just rotation about the origin at a rate $\Omega_\textsc{p}$. Combining these all together we have that the Gaussian version of the update map \mbox{$\Phi_\text{cell}^\text{S} = \mathcal{U}_0^2\circ \Phi^\text{I}_2 \circ \Phi^\text{I}_1$} is,
\begin{align}\label{GaussianScell}
\Phi^\text{S}_\text{cell}:& \ \ \ 
\sigma_\textsc{p} \to \sigma_\textsc{p}'''=T^\text{S}_\text{cell} \ \sigma_\textsc{p} \ T_\text{cell}^\text{S}{}^\intercal +R_\text{cell}^\text{S},
\end{align}
where
\begin{align}
\Phi^\text{I}_1:\quad\sigma_\textsc{p} &\to
\sigma_\textsc{p}'= T_1^\text{I} \ \sigma_\textsc{p} \ T_1^\text{I}{}^\intercal +R_1^\text{I},\\
\Phi^\text{I}_2: \quad
\sigma_\textsc{p}' &\to\sigma_\textsc{p}'' =T_2^\text{I} \ \sigma_\textsc{p}' \ T_2^\text{I}{}^\intercal +R_2^\text{I},\\
\mathcal{U}_0^2:\quad
\sigma_\textsc{p}'' &\to
\sigma_\textsc{p}'''=\mathcal{R}(2\,\Omega_\textsc{p}\,\tau_\text{max}) \ \sigma_\textsc{p}'' \ \mathcal{R}(2\,\Omega_\textsc{p}\,\tau_\text{max})^\intercal.
\end{align}

\subsection{Gaussian Interpolated Collision Model formalism}
Now that we have discussed how $\Phi^\text{S}_\text{cell}$ can be efficiently computed we need a way to analyze the effect of repeated application of this map. Our immediate thought may be to find the eigendecomposition for $\Phi^\text{S}_\text{cell}$ in order to figure out its fixed points and convergence rates. This approach is complicated by the fact that our update map 1) acts on a matrix and 2) is linear-affine not linear.

These difficulties can be overcome by the following two isomorphisms. The first isomorphism is the vectorization map, $\vec$, which maps outer products to tensor products as $\vec(\bm{u}\bm{v}^\intercal)=\bm{u}\otimes\bm{v}$. By linearity this defines the map's action on all matrices. Note that this map has the property that $\vec(A\,B\,C^\intercal)=A\otimes C \, \vec(B)$. Applying this map to our Gaussian update equation \eqref{GaussianScell} we find,
\begin{align}\label{VecUpdate}
\Phi^\text{S}_\text{cell}:& \ \ \ 
\vec(\sigma_\textsc{p}) \to T_\text{cell}^\text{S}\otimes T_\text{cell}^\text{S} \ \vec(\sigma_\textsc{p}) + \vec(R_\text{cell}^\text{S}).    
\end{align}
The second isomorphism we apply is embedding the vec operation into an affine space as, $\vec(\sigma_\textsc{p})\leftrightarrow(1,\vec(\sigma_\textsc{p}))$. Using this we can rewrite \eqref{VecUpdate} as,
\begin{align}
\Phi^\text{S}_\text{cell}:\quad
\begin{pmatrix}
1\\
\vec(\sigma_\textsc{p})
\end{pmatrix}
\to\begin{pmatrix}
1 & 0\\
\vec(R_\text{cell}^\text{S}) & T_\text{cell}^\text{S} \otimes T_\text{cell}^\text{S}
\end{pmatrix}
\begin{pmatrix}
1\\
\vec(\sigma_\textsc{p})
\end{pmatrix}
=M_\text{cell}^\text{S}
\begin{pmatrix}
1\\
\vec(\sigma_\textsc{p})
\end{pmatrix}.
\end{align}
We can now analyze the dynamics generated by repeated application of $\Phi^\text{S}_\text{cell}$ by studying  $M_\text{cell}^\text{S}$. In particular we will study $M_\text{cell}^\text{S}$ in two ways, 1) by computing its eigenvectors and eigenvalues and 2) by computing its logarithm. Note that $M_\text{cell}^\text{S}$ is a $5\times5$ real matrix and so both of these tasks can be done easily.

If $M_\text{cell}^\text{S}$ has a unique eigenvector, $\bm{v}_{\lambda=1}$, with eigenvalue $\lambda=1$ then $M_\text{cell}^\text{S}$ has a one-dimensional fixed-point space. Moreover, if all other $\lambda<1$ then this fixed-point space is attractive. Our simulations show that for all parameters under consideration both of these conditions hold. 

This in turn implies that repeated applications of $\Phi^\text{S}_\text{cell}$ to any $\sigma_\textsc{p}(0)$ will drive the state to a unique attractive fixed point, $\sigma_\textsc{p}(\infty)$. To see this,  note that our states lie on an affine subspace, i.e. $\bm{v}=(1,\vec(\sigma_\textsc{p}))$. This affine subspace will intersect the 1D fixed-point space of $M_\text{cell}^\text{S}$ exactly once. Concretely, normalizing $\bm{v}_{\lambda=1}$ to lie in the affine subspace (i.e., such that its first component is one) we have $\bm{v}_{\lambda=1}=(1,\vec(\sigma_\textsc{p}(\infty)))$.

We can analyze the other eigenvectors and eigenvalues to get an idea of how this fixed point is approached (i.e., from which directions at which rates).  That is, we can study the decoherence modes and decoherence rates. However, direct examination of the eigenvectors proves unilluminating.  To more clearly identify the dynamics' decoherence modes, we can make use of the ICM formalism~\cite{GrimmerThesis,RRI,Pure}, particularly in its Gaussian form~\cite{GAB}. 

Roughly speaking, the ICM formalism takes a given discrete-time repeated-update  dynamics and constructs the unique Markovian and time-independent differential equation which interpolates between the discrete time points, with no approximation at the points between which we interpolate. In our case the discrete dynamics,
\begin{align}
\begin{pmatrix}
1\\
\vec(\sigma_\textsc{p}(n\,\delta t))
\end{pmatrix}
=\big(M_\text{cell}^\text{S}{}\big)^n
\begin{pmatrix}
1\\
\vec(\sigma_\textsc{p}(0))
\end{pmatrix},
\end{align}
can be interpolated by the differential equation,
\begin{align}
\frac{\d}{\d t}
\begin{pmatrix}
1\\
\vec(\sigma_\textsc{p}(t))
\end{pmatrix}
=\mathcal{L}
\begin{pmatrix}
1\\
\vec(\sigma_\textsc{p}(t))
\end{pmatrix},
\end{align}
where $\mathcal{L}=\frac{1}{\delta t} \text{Log}(M_\text{cell}^\text{S})$. One can easily check that this interpolation exactly matches the discrete update at every $t=n\,\delta t$. From this interpolation scheme we can isolate the dynamics of the covariance matrix, $\sigma_\textsc{p}(t)$. After some work~\cite{GAB} one finds a master equation for $\sigma_\textsc{p}(t)$ of the form,
\begin{align}
\frac{\d}{\d t}\sigma_\textsc{p}(t) 
&=(\Omega A) \ \sigma_\textsc{p}(t)
+ \sigma_\textsc{p}(t) \ (\Omega A)^\intercal
+C,
\end{align}
where,
\begin{align}
\Omega A&=\frac{1}{\delta t}\text{Log}(T_\text{cell}^\text{S}),\\
C&=\frac{1}{\delta t}\frac{\text{Log}(T_\text{cell}^\text{S}\otimes T_\text{cell}^\text{S})}{T_\text{cell}^\text{S}-\openone} \ \vec(R_\text{cell}^\text{S}).
\end{align}
This Gaussian master equation can then be analyzed in terms of its decoherence rates and decoherence modes in a standard way~\cite{GaussianClass}. For instance, $C$ can be understood as a noise term and $A$ can be broken down into rotation, squeezing and relaxation effects.

\section{Characterizing Temperature and Thermality of the Final Detector State}\label{AppSec3}

\begin{figure}[ht]
	\centering
	\includegraphics[width=0.45\linewidth]{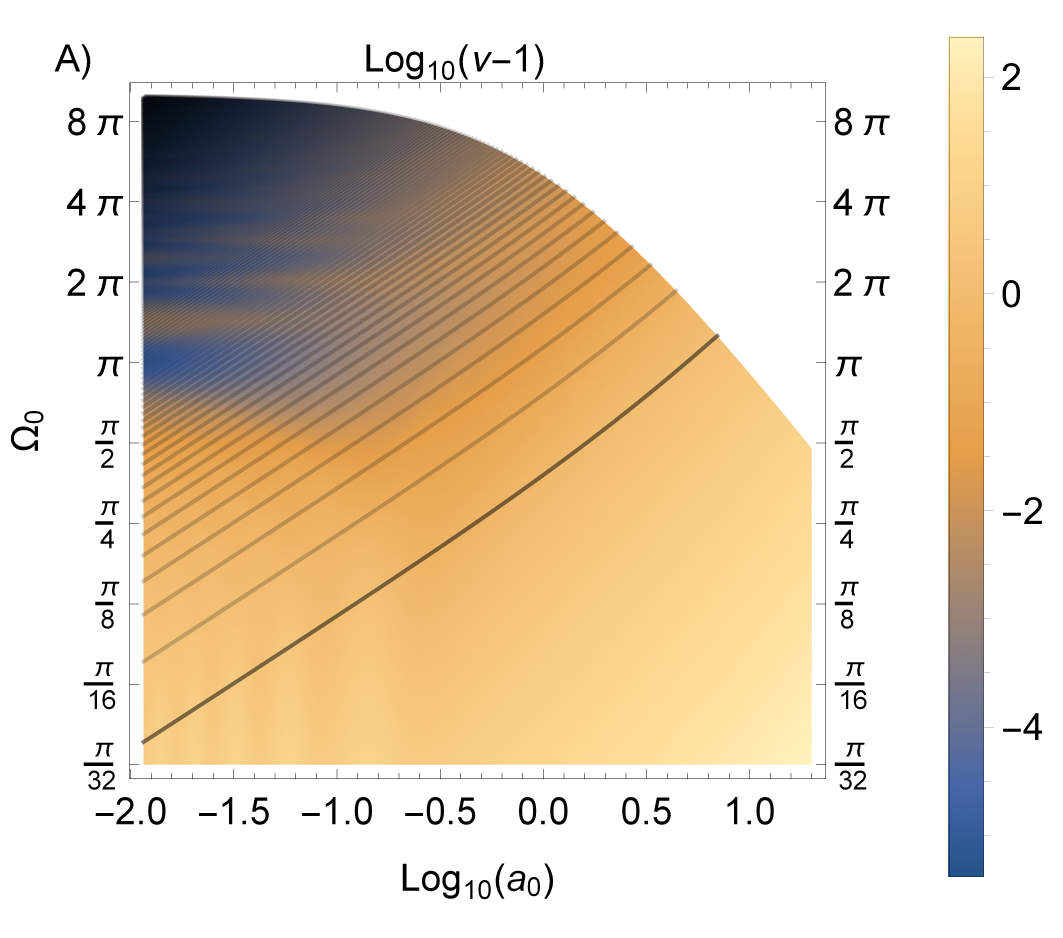}
	\includegraphics[width=0.45\linewidth]{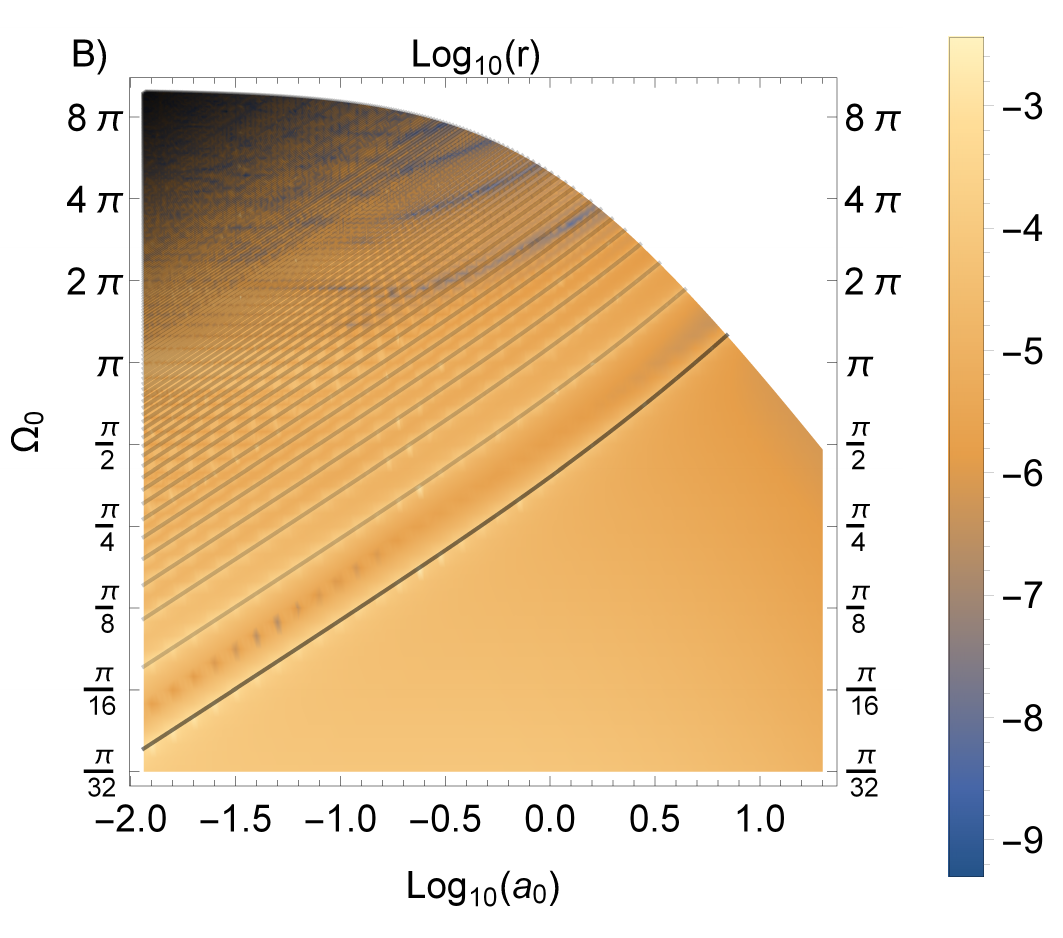}
	\caption{The symplectic eigenvalue $\nu$ and the squeezing parameter, $r$, of the final probe state $\sigma_\textsc{p}(\infty)$ are shown in A) and B) respectively. Note that the axes are all on a logarithmic scale.}
	\label{SuppNuAndR}
\end{figure}

As we have discussed in the main text, we can efficiently compute the final covariance matrix of the detector, $\sigma_\textsc{p}(\infty)$, after it has traveled through many cells. To characterize this state we can write it in the standard form,
\begin{align}
\sigma_\textsc{p}(\infty)
=\mathcal{R}(\theta)\begin{pmatrix}
\nu \exp(r) & 0 \\
0 & \nu \exp(-r) \\
\end{pmatrix}\mathcal{R}(\theta)^\intercal
\end{align}
for some symplectic eigenvalue $\nu\geq1$, squeezing parameter $r>0$ and angle $\theta\in[-\pi/2,\pi/2]$ where $\mathcal{R}(\theta)$ is the $2\times2$ rotation matrix. The values of $\nu$ and $r$ are shown in Figure \ref{SuppNuAndR} as functions of $a_0=a L/c^2$ and $\Omega_0=\Omega_\textsc{p} L/c$. Note that $r\lesssim 10^{-3}$ whereas $\nu-1\lesssim 10^2$. Thus it appears that for the range of parameters we consider the final state of the detector is not very squeezed and is therefore approximately thermal. But how can we quantify the degree to which the state is thermal?

In this section we will establish that this state is in fact approximately thermal by showing that $r$ is ``small'' in several different ways.  Moreover, we will also explain the interesting band-like structure which appears in the plot of the squeezing parameter.

\subsection{Thermality Criteria}
Let us first consider the method of assessing thermality mentioned in the main text, and originally introduced in~\cite{BrennaTherm}. Namely, we quantify how the energy needed to build the state from the vacuum is divided between the energy spent on squeezing and the energy spent on heating it to the corresponding unsqueezed thermal state. Concretely, the ratio of these energies is given by the following expression,
\begin{align}
&\delta(\nu,r)=\left\vert\frac{ E(\nu,r)-E(\nu,0)}{E(\nu,0)}\right\vert
=\frac{\nu(\text{cosh}(r)-1)}{\nu-1}
=\frac{\nu\,r^2}{\nu-1}+O(r^4),
\end{align}
where $E(\nu,r)=\hbar\Omega_\textsc{p}\left(\nu\,\text{cosh}(r)-1\right)$ is the average energy of a generic squeezed-thermal state. Note that the ground state (with $\nu=1$ and $r=0$) has (by convention) zero energy. We can use $\delta$ as a thermality criterion: if $\delta\ll 1$ then the state's squeezing energy is much less than its thermal energy. Note that the $\delta$ test is harder to pass the nearer we are to the ground state. That is, for fixed $r>0$ we have $\delta$ diverging as $\nu\to1$.

\begin{figure}[ht]
	\centering
	\includegraphics[width=0.45\linewidth]{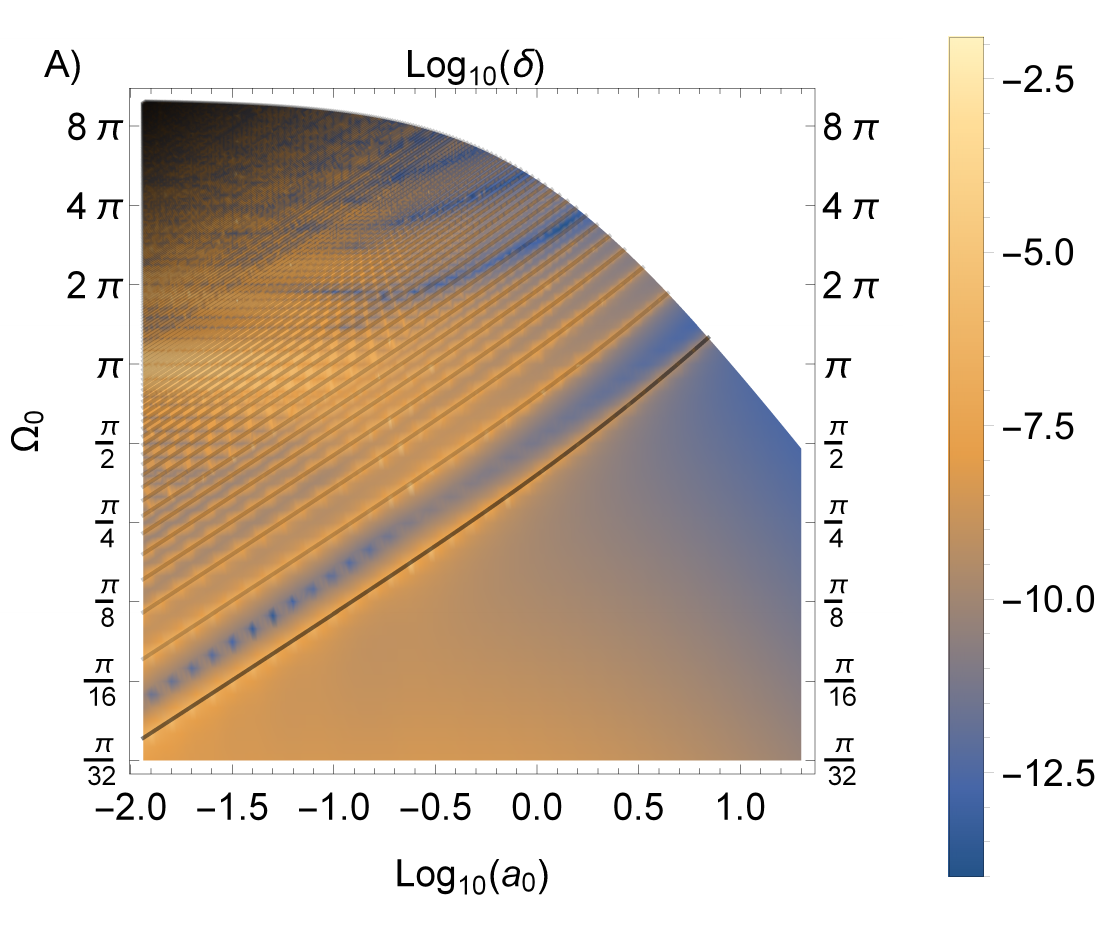}
	\includegraphics[width=0.45\linewidth]{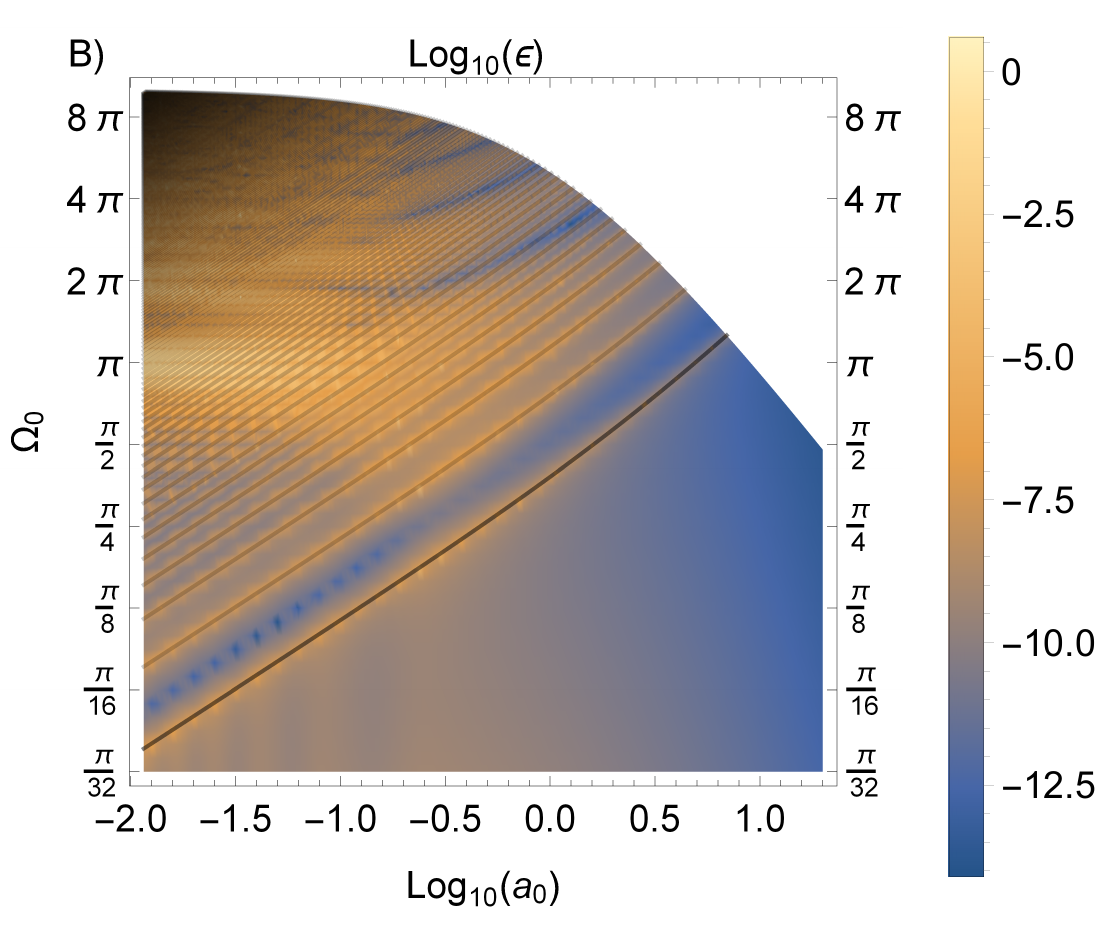}
	\caption{The thermality measures $\delta$ and $\epsilon$ of the final probe state $\sigma_\textsc{p}(\infty)$ are shown in A) and B) respectively. Note that the axes are all on a logarithmic scale.}
	\label{SuppDAndE}
\end{figure}

Figure~\ref{SuppDAndE}A) shows that $\delta\lesssim 10^{-5}$ in the regime where we see the Unruh effect. Thus the state can be deemed very nearly thermal by this measure.

Another approach to characterizing the thermality of a Gaussian state is to generate a few different temperature estimates and demand their relative differences be small. A series of temperature estimates can be found by considering the relative populations of the detector's energy levels. The probability of measuring a generic single-mode squeezed-thermal state, $\sigma_\textsc{p}(\nu,r)$, and finding $n$ excitations is,
\begin{align}\label{SuppPs}
	P_{n=0}(\nu,r) &= \frac{2}{[(1+\lambda_1)(1+\lambda_2)]^{1/2}}, \\
	P_{n=1}(\nu,r) &= \frac{2(\lambda_1 \lambda_2 - 1)}{[(1+\lambda_1)(1+\lambda_2)]^{3/2}}, \\
	P_{n=2}(\nu,r) &=
	\frac{2 + \lambda_1^2 + \lambda_2^2 - 6 \lambda_1 \lambda_2 + 2 \lambda_1^2 \lambda_2^2}{[(1+\lambda_1)(1+\lambda_2)]^{5/2}},
\end{align}
where $\lambda_1=\nu\exp(r)$ and $\lambda_2=\nu\exp(-r)$ are the eigenvalues of $\sigma_\textsc{p}(\nu,r)$. These expressions can be calculated straightforwardly by taking the overlap of a generic Gaussian Wigner function with the Fock state Wigner functions. From these we can compute the excitation de-excitation ratio (EDR) temperature estimates as,
\begin{align}
k_\textsc{b}\,T_{nm}^\text{EDR}&=\frac{(m-n)\,\hbar\Omega_\textsc{p}}{\text{ln}(P_n/P_m)}.
\end{align}
We can declare that a state is reasonably thermal if many of its EDR temperature estimates between different energy levels all agree. For instance, we may consider the relative difference,
\begin{align}
\left\vert\frac{T_{02}^\text{EDR}-T_{01}^\text{EDR}}{T_{01}^\text{EDR}}\right\vert\ll1.
\end{align}
Expanding this relative difference for small $r$ we find,
\begin{align}
\left\vert\frac{T_{02}^\text{EDR}-T_{01}^\text{EDR}}{T_{01}^\text{EDR}}\right\vert
&=\epsilon(\nu,r)+O(r^4);\qquad
\epsilon(\nu,r)
=\frac{\nu^2\, r^2}{2(\nu^2-1)^2 \text{arccoth}(\nu)}.
\end{align}
We can take $\epsilon\ll1$ to be an alternate thermality criterion to $\delta\ll1$. Contrasting $\delta$ and $\epsilon$ we can see that $\epsilon$ is a harder test to pass, especially for near-ground states. That is, for fixed $r>0$, we have that $\epsilon$ diverges faster than $\delta$ as $\nu\to1$.

Figure~\ref{SuppDAndE}B) shows that $\epsilon$ over the range of parameters we consider. Despite $\epsilon$ being a harder test, we still find that the final probe state is approximately thermal (with respect to $\epsilon$), at least in the regime where we see the Unruh effect. Namely, in the lower-right region of the plot we have $\epsilon\lesssim10^{-5}$.

In addition to $\delta$ and $\epsilon$ we have considered several other thermality measures, including comparing the EDR temperature estimates between different levels (e.g., $T_{12}^\text{EDR}$ versus $T_{01}^\text{EDR}$) as well as more information-theoretic measures (e.g., Hellinger and total variation distances). In each case these measures have indicated that the probe state is effectively indistinguishable from thermal in the regime where we see the Unruh effect.

\subsection{Explaining the bands}
Looking at Figure~\ref{SuppNuAndR}B) one may notice that there are bands of increased squeezing appearing in an ordered way. (The corresponding bands in Figure~\ref{SuppDAndE} are a consequence of this increased squeezing). We will now explain why these appear and why they are where they are.

The relevant quantity is the phase that the probe operators rotate through as the probe crosses one cavity, \mbox{$\Theta=\Omega_\textsc{p}\tau_\text{max}$}. Indeed, the bands lie on (or very near to) the $\Theta=n\pi/2$ lines shown in Figures \ref{SuppNuAndR} and \ref{SuppDAndE}. Note that the $\Theta=\pi/2$ line is bold.

We can explain the occurrence of these bands as follows. Recall that the update map which we repeatedly apply is $\Phi_\text{cell}^\text{S}=\mathcal{U}_0^2\circ\Phi^\text{I}_2\circ\Phi^\text{I}_1$. Recall further that, in the interaction picture, the update map for crossing the first cell is $\Phi^\text{I}_{1,2}=\Phi^\text{I}_2\circ\Phi^\text{I}_1$. Suppose that the effect of $\Phi^\text{I}_{1,2}$ is to squeeze the state in some direction $\theta_\text{squ}(a_0,\Omega_0)$ and then rotate it by an amount $\theta_\text{rot}(a_0,\Omega_0)$. The effect of $\Phi_\text{cell}^\text{S}$ would then be to squeeze the state in some direction $\theta_\text{squ}(a_0,\Omega_0)$ and then rotate it by an amount $\theta_\text{rot}(a_0,\Omega_0)+2\Theta$.

First let us analyze the case where the effect of $\Phi_\text{cell}^\text{S}$ is a quarter-turn,  $\theta_\text{rot}(a_0,\Omega_0)+2\Theta=\pi/2$. In this case the second application of $\Phi_\text{cell}^\text{S}$ would immediately undo the squeezing done by the first application of $\Phi_\text{cell}^\text{S}$. A similar phenomenon will happen for most values of $\theta_\text{rot}(a_0,\Omega_0)+2\Theta$. Over many applications of $\Phi_\text{cell}^\text{S}$ the state will have been squeezed in every direction more-or-less equally. The result in this case would be a minimally squeezed state. 

The exception to this argument is when $\theta_\text{rot}(a_0,\Omega_0)+2\Theta= n \pi$. In this case the state is left unchanged by the rotation (note that squeezed states have a $\pi$-rotational symmetry) such that it is squeezed in the same direction every time. This squeezing does not become infinite, however, as the dynamics also includes a relaxation rate. Thus we expect a spike in the squeezing of the final state when $\Theta=n\pi/2-\theta_\text{rot}(a_0,\Omega_0)/2$.

Finally, we note that we have reason to believe that $\theta_\text{rot}(a_0,\Omega_0)$ is small for all $a_0$ and $\Omega_0$. Recall that $\theta_\text{rot}(a_0,\Omega_0)$ is the amount of rotation given by the interaction picture map $\Phi^\text{I}_{1,2}$. The interaction picture is designed to remove the free evolution/rotation of the system. Thus $\theta_\text{rot}(a_0,\Omega_0)$ only corresponds to the rotation induced in the probe by the interaction Hamiltonian. Thus we expect spikes in the squeezing at $\Theta\approx n\pi/2$ which is just what we see.

\section{Details on Mode Convergence}\label{ModeConvergence}
As we discussed in the main text, we truncate the number of cavity modes considered in order to make our computations tractable. In this section we study the convergence of our results with the number of cavity modes considered.

We expect our scenario to have better convergence behaviour than other previous studies on probes accelerating inside optical cavities (such as e.g., \cite{BrennaTherm}) since in our setup the probe does not reach ultrarelativistic speeds with respect to the cavity walls. As such, the probe's gap $\Omega_\textsc{p}$ does not sweep across many cavity modes as it is blue/red-shifted ($\Omega_\textsc{p}\leftrightarrow\gamma_\text{max} \Omega_\textsc{p}$) with respect to the lab frame. For instance, with $\Omega_0=\pi/16$ and $a_0=10$ we have $\gamma_\text{max}=1+a_0$ such that $\gamma_\text{max} \Omega_0=11\pi/16$. Note that even when maximally blue shifted, the probe frequency is still below the frequency of the first cavity mode $\omega_0=\pi$.

Another reason that one may worry that many cavity modes are required for convergence is that the probe suddenly couples/decouples from each cavity. Indeed, one can think of the probe having a top-hat switching function, $\chi(\tau)$. In general one would expect that such a sudden change in the coupling would make high frequency cavity modes relevant. However, a key design feature of our setup regulates the suddenness of this switching. Namely, the cavity's Dirichlet boundary conditions enforce that the probe is effectively decoupled from the field at the time of this switching.

\begin{figure}[ht]
	\centering
	\includegraphics[width=0.75\linewidth]{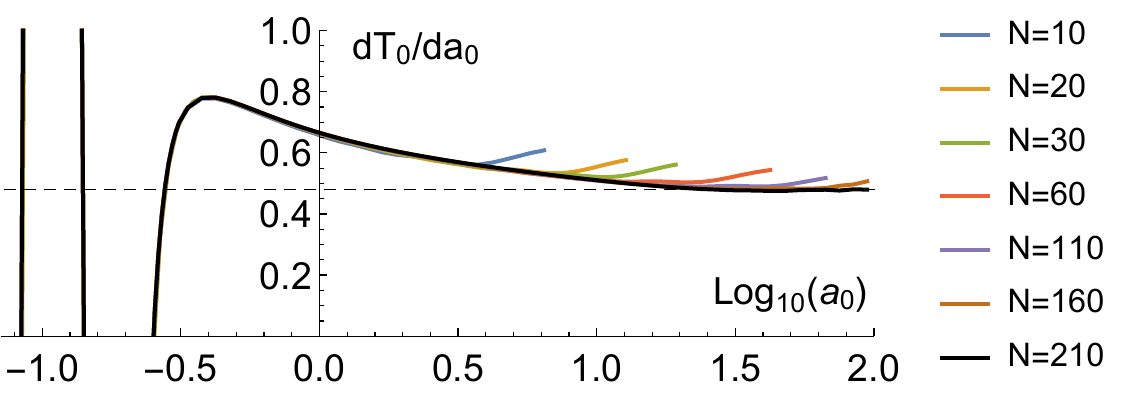}
	\caption{Derivative of the  probe's final dimensionless temperature \mbox{$T_0=k_\textsc{b} T   L/\hbar c$} with respect to the  acceleration \mbox{$a_0= a L/c^2$} as a function of $a_0$ on log-scale. The dimensionless probe gap, \mbox{$\Omega_0=\Omega_\textsc{p} L/c=\pi/16$}, and the dimensionless coupling strength, $\lambda_0=\lambda L/\sqrt{\hbar c}=0.01$, are fixed. The black-dashed line is at $\d T_0/\d a_0=1/2$. The colored lines show the values of $\d T_0/\d a_0$ which result from considering only $N$ cavity modes where $N=10$, $20$,  $30$,  $60$,  $110$,  $160$, and $210$. These lines split off from the rest one at a time in order from left to right.}
	\label{Convergence}
\end{figure}

Taken together, these suggest that not too many cavity modes will be needed for convergence. Let's see how these expectations play out when we actually put them to the test. Fig.~\ref{Convergence} shows the $\Omega_0=\pi/16$ line of Fig.~1b) of the main text converging as we increase the number of field modes, $N$, which we consider. Unsurprisingly, as the acceleration increases we require more cavity modes for convergence. Fig.~\ref{Convergence} suggests that using $N=20$ modes is sufficient when $a_0\lesssim6$ and that using $N=200$ is sufficient when $a_0\lesssim 100$.

\twocolumngrid

\bibliography{references}

\end{document}